\documentstyle[aaspp4,12pt]{article}

\def\simlt{\lower.5ex\hbox{$\; \buildrel < \over \sim \;$}}
\def\simgt{\lower.5ex\hbox{$\; \buildrel > \over \sim \;$}}
\def\gsim{\;\rlap{\lower 2.5pt
\hbox{$\sim$}}\raise 1.5pt\hbox{$>$}\;}
\def\lsim{\;\rlap{\lower 2.5pt
   \hbox{$\sim$}}\raise 1.5pt\hbox{$<$}\;}

\def\spose#1{\hbox to 0pt{#1\hss}}
\def\lta{\mathrel{\spose{\lower 3pt\hbox{$\mathchar''218$}}
     \raise 2.0pt\hbox{$\mathchar''13C$}}}
\def\gta{\mathrel{\spose{\lower 3pt\hbox{$\mathchar''218$}}
     \raise 2.0pt\hbox{$\mathchar''13E$}}}
\newcommand{\beq}{\begin{equation}}
\newcommand{\eeq}{\end{equation}}


\begin{document}

\title{On The Dark Side of Quasar Evolution}

\author{Kristen Menou\altaffilmark{a} \& Zolt\'an Haiman\altaffilmark{b}}

\affil{$^{a}$ Institut d'Astrophysique de Paris, 98bis Boulevard Arago, 
75014 Paris, France}
\affil{$^{b}$ Department of Astronomy, Columbia University, 
550 West 120th Street, New York, NY 10027, USA}

\authoremail{menou@iap.fr}
\authoremail{zoltan@astro.columbia.edu}


\begin{abstract}

Recent improved determinations of the mass density $\rho_{\rm BH}$ of
supermassive black holes (SMBHs) in the local universe have allowed
accurate comparisons of $\rho_{\rm BH}$ with the amount of light
received from past quasar activity.  These comparisons support the
notion that local SMBHs are ``dead quasars'' and yield a value
$\epsilon \gsim 0.1$ for the average radiative efficiency of cosmic
SMBH accretion.  BH coalescences may represent an important component
of the quasar mass assembly and yet not produce any observable
electromagnetic signature. Therefore, ignoring gravitational wave (GW)
emission during such coalescences, which reduces the amount of mass
locked into remnant BHs, results in an overestimate of $\epsilon$.
Here, we put constraints on the magnitude of this bias.  We calculate
the cumulative mass loss to GWs experienced by a representative
population of BHs during repeated cosmological mergers, using loss
prescriptions based on detailed general relativistic calculations. 
Despite the possibly large number of mergers in the assembly 
history of each individual SMBH,
we find that near--equal mass mergers are rare, and therefore
the cumulative loss is likely to be modest, amounting at
most to an increase by 20 percent of the inferred $\epsilon$
value. Thus, recent estimates of $\epsilon \gsim 0.1$ appear robust.
The space interferometer {\it LISA} should provide empirical
constraints on the dark side of quasar evolution, by measuring the
masses and rates of coalescence of massive BHs to cosmological
distances.

\end{abstract}


\section{Introduction}

It is now widely accepted that quasar activity is powered by accretion
onto supermassive black holes (SMBHs). From the active phases of
accretion which characterize luminous, high-redshift quasars, one
expects remnant SMBHs to be present at the centers of nearby galaxies
(Lynden-Bell 1969; Soltan 1982; Rees 1990). The evidence for such a
population of dead quasars has been growing over the years (see
Kormendy \& Richstone 1995 for a review) and it is now compelling
(Magorrian et al. 1998).

Dynamical studies of nearby massive galaxies indicate that a close
link exists between the masses of dead quasar SMBHs and the properties
of their host galaxies, including the spheroid's mass (Magorrian et
al. 1998; Haering \& Rix 2004), velocity dispersion (Ferrarese \&
Merritt 2000; Gebhardt et al. 2000; Tremaine et al. 2002) and the
total galactic mass (Ferrarese 2002). These empirically-determined
correlations allow accurate tests of the idea that the amount of mass
locked into SMBHs in nearby dead quasars should be comparable to that
inferred from the amount of light received from past quasar activity,
with a radiative efficiency $\epsilon \sim 10\%$, since the latter is
a tracer of BH mass build up via accretion (Soltan 1982; Chokshi \&
Turner 1992). Recent comparisons do find a good agreement between the
mass density in dead quasar SMBHs and the integrated light from
optically-bright quasars, provided that the radiative efficiency of BH
accretion in luminous quasars is $\epsilon \gsim 0.1$ (Yu \& Tremaine
2002; Aller \& Richstone 2002; Haiman, Ciotti \& Ostriker 2004). 
{The luminosity density of the quasar population can also be inferred
from the X-ray bands.  This has led to suggestions that optical quasar
surveys may be missing some of the quasar emission (because of
obscuration; Fabian \& Iwasawa 1999; Barger et al. 2001), which may be
indicative of radiatively more efficient accretion onto fast-spinning
BHs (Elvis, Risaliti \& Zamorani 2002).  However, recent work, using
the soft X-ray luminosity function of Miyaji et al. (2001) have found
a low efficiency of $\epsilon\sim 0.05$ (Haiman, Ciotti \& Ostriker
2004).  Soft X-ray bands miss the most highly obscured sources, but
the efficiency is increased further only by a factor of $\sim$ two
when hard X-ray sources (with the luminosity function from Ueda et
al. 2003) are added in the comparison (Marconi et al. 2003). }

In the present study, we investigate the possibility that cumulative
mass-energy losses to gravitational waves (GWs) during repeated BH
binary coalescences, in the context of standard cosmological
hierarchical structure formation models, may significantly reduce the
amount of mass currently locked into BHs, and thus effectively bias
the comparison between active and dead quasars toward larger values of
the radiative efficiency, $\epsilon$. The role of GW losses for the
quasar population has already been considered by Yu \& Tremaine
(2002), but only with an idealized description of cosmological mergers
and for maximized ``adiabatic'' losses to GWs (see also Ciotti \& van
Albada 2001; Volonteri et al. 2003; Koushiappas et al. 2004).  In a companion
paper (Menou \& Haiman 2004; hereafter paper I), we have reconsidered
this issue with a more realistic description of cosmological BH
mergers. Our results suggested that, while the mass loss in a single merger
event is small, after numerous repeated mergers over cosmic times, adiabatic
losses can result in a substantial and astrophysically important
reduction of the BH mass density. However, the adiabatic assumption
provides only an upper bound on the mass-energy carried away by GWs, and
thus largely overestimates realistic losses. Here, we use an improved
prescription for GW losses, based on the latest available general
relativistic calculations, to provide more accurate constraints on the
possible role of GWs in modifying the mass budget of merging quasars.

\section{Models}

\subsection{Merger History of Massive Black Holes}

Our description of the cosmological merger history of massive BHs
follows very closely that presented in paper I (see also Menou, Haiman
\& Narayanan 2001 for details). We use a dark matter halo merger tree
with a standard $\Lambda$CDM cosmology to evolve the population of
massive BHs and their host galaxies. Only galaxies with a total mass
exceeding a virial temperature equivalent of $10^4$~K are described by
the tree, since these are the galaxies with efficient enough baryon
cooling to allow BH formation (smaller objects rely on ${\rm H_2}$
cooling and are subject to disruptive feedback; Oh \& Haiman 2004). It
is assumed by default in our models that all potential host galaxies
do harbor a massive BH, although we have also explored models in which
massive BHs are ten times rarer and are initially confined to the $10 \%$
most massive galaxies (as described in paper I).

Recent quasar evolutionary studies indicate that the majority of the
mass currently locked into SMBHs was accreted between redshifts $z
\simeq 3$ and $z=0$ (e.g. Yu \& Tremaine 2002; Marconi et
al. 2004). Since most of the losses due to mergers is expected
to occur when most of the BH mass is being assembled, 
in order to estimate the GW losses due to mergers, there is no
need to extend our models much beyond redshifts
$z \sim 3$.
We assume that the same relation between SMBHs and their host galaxy
velocity dispersion exists at $z \simeq 3$ as it does locally (Shields
et al. 2003) and adopt a mass ratio between BHs and their parent
halos given by (Ferrarese 2002; Wyithe \& Loeb 2004):
\begin{equation} 
M_{\rm bh}= 10^9 M_\odot \left( \frac{M_{\rm halo}}{1.5 \times 10^{12}
M_\odot}\right)^{5/3} \left( \frac{1+z}{7} \right)^{5/2},
\end{equation}
where $M_{\rm halo}$ is the mass of the dark matter halo associated
with each galaxy.  This relation may result from the BH mass being
limited (at least initially, during the luminous quasar phase) by
feedback from the quasar's radiation (Silk \& Rees 1998; Wyithe \&
Loeb 2003). We have found in exploratory models (see paper I) that the
shape of this $M_{\rm bh}-M_{\rm halo}$ relation is not strongly
modified by the redistribution of BHs in galaxies due to cosmological
mergers. This provides additional motivation for setting up BH masses
according to equation~(1) at $z \simeq 3$ and neglecting the role that
accretion may subsequently have in modifying them over cosmic times
(modulo an overall scaling factor). Below, we will express all mass
deficits in evolutionary models with GW losses relative to a no-loss
model, thus effectively scaling out the exact $\rho_{\rm BH}$ value
from the loss problem.

Starting at $z \simeq 3$, we let the BH population evolve through a
series of cosmological mergers up until $z=0$. We assume that each
time two galaxies merge, the two BHs they were hosting coalesce
rapidly. {In doing so, we ignore complications related to the
``last parsec'' problem for BH coalescences (Begelman, Blandford \&
Rees 1980; see Milosavljevic \& Merritt 2003 for a recent discussion)
and effectively maximize BH merger rates in our models. Rapid
coalescences may be induced by effects due to the triaxility of
galaxies (Yu 2002) or the presence of surrounding gas (Gould \& Rix
2000; Armitage \& Natarajan 2002; Escala et al. 2004). We do account
for the inefficiency of dynamical friction in initially bringing the
two BHs together, however,} by assuming, following Yu (2002), that BH
binaries do not form for mass ratios $q < 10^{-3}$ (see also paper I).
Finally, we emphasize that the above model, which ignores gas
accretion, is not intended to yield a realistic description of the
quasar BH population. Rather, our limited goal here is to quantify the
effect of mergers alone on the remnant BH mass budget.

\subsection{Mass Loss to Gravitational Waves}

\clearpage

\begin{table*}
\caption{GRAVITATIONAL WAVE LOSS PRESCRIPTIONS}
\begin{center}
\begin{tabular}{cccc} \hline \hline
\\ BH Spin Limit & Inspiral& Plunge & Ringdown \\ (1)&(2)&(3)&(4)\\ \\
\hline \\ Slow Spin & $\alpha_{\rm ins}=0.06$ & $\alpha_{\rm
plu}=0.01$ & $\alpha_{\rm rin}=10^{-5}$\\ Fast Spin & $\alpha_{\rm
ins}=0.42$ & $\alpha_{\rm plu}=0.10$ & $\alpha_{\rm rin}=0.03$\\ \\
\hline
\end{tabular}
\label{tab:one}
\end{center}
\end{table*}

\clearpage

In their final stages of coalescence, energy and momentum are
extracted from massive BH binaries by emission of GWs. As a result,
the BH merger remnant has a mass which is less than that of its two
progenitors. This mass loss to GWs, and its cumulative effect on the
global BH mass density through repeated cosmological mergers, is the
main focus of our study.

Rather than adopting simple GW loss prescriptions as in paper I, we
wish to obtain more accurate constraints based on the latest available
general relativistic calculations. This is no easy task, however,
because the general relativistic BH coalescence problem has not been
solved in full generality (see Baumgarte \& Shapiro 2003 for a review
of numerical progress) and approximate analytical solutions exist only
for some limiting cases. Here, we will use such approximate solutions
and extrapolate them whenever necessary.

Let us denote by $m$ and $M$ the masses of the two BHs involved in a
coalescence, with $m \leq M$. The mass ratio is $q=m/M \leq 1$ and the
reduced mass is defined as $\mu=mM/(m+M)$.  In the test particle limit
($q \ll 1$), the coalescence can be decomposed into three successive
phases: (i) a slow inspiral phase during which the two BHs spiral in
quasi-adiabatically on nearly circular orbits, (ii) a plunge phase due
to the existence of an innermost stable circular orbit (ISCO) past which
the two BHs are brought together via a dynamical instability, and
(iii) a final ringdown phase during which the perturbed merger remnant
relaxes to a stationary Kerr BH. The separation between the plunge and
ringdown phases is somewhat arbitrary. In addition, it is possible
that no ISCO exists for some combinations of BH masses and spins when
$q \rightarrow 1$. Clearly then, the decomposition into three successive
phases must be used with caution. It is useful, however, in that
approximate solutions for GW losses have been derived in some limiting
cases for some of these phases.

We consider the two extreme limits for the spins of BHs involved in
coalescences. In the slow-spin limit, BHs are assumed to have no
spin. In the fast-spin limit, BHs are assumed to be maximally rotating
(with a spin parameter $a \equiv J_{\rm bh}/M_{\rm bh}=1$, in $c=G=1$
units).  In a given evolutionary model, we assume for simplicity that
all the BHs involved satisfy one or the other spin limits. If we
generically write a mass loss from the BH binary as $\Delta(m+M)$,
then the losses that we have adopted in our models for the inspiral,
plunge and ringdown phases are, respectively,
\begin{eqnarray}
\Delta(m+M)_{\rm ins} & = &- \alpha_{\rm ins} \mu,\\
\Delta(m+M)_{\rm plu} & = &- \alpha_{\rm plu} M q^2,\\
\Delta(m+M)_{\rm rin} & = &- \alpha_{\rm rin} M_{coa} q^2,
\end{eqnarray}
where the ``coalesced'' mass (before ringdown starts) is $M_{\rm coa}
= m+M-\Delta(m+M)_{\rm ins}-\Delta(m+M)_{\rm plu}$. The values of the
loss coefficients $\alpha$ are given in Table~\ref{tab:one} for the
two spin limits. Justifications for these prescriptions follow.

Losses to GWs during the inspiral phase have been discussed
extensively. They involve calculating the binding energy at the
location of the ISCO, since the quasi-adiabatic inspiral experienced
by the binary means an efficient loss of this binding energy to GWs
via a succession of nearly circular orbits. In the test particle limit
($m \ll M$), it is well known that the loss during inspiral is $\sim 6
\%$ of $mc^2$ in the slow spin limit, and $\sim 42 \%$ of $mc^2$ in
the fast (prograde) spin limit (as is the case for accretion
efficiency; see, e.g., Shapiro \& Teukolsky 1983). In the equal-mass
binary limit ($m = M$), results have been derived under a variety of
approximations. For non-spinning, equal-mass BHs, the binding energy
per unit reduced mass at the ISCO is roughly consistent with the test
particle result (see Table~1 in Gammie, Shapiro \& McKinney 2004). For
spinning, equal-mass BHs, the analysis of Pfeiffer, Teukolsky \& Cook
(2000; their Table~1) indicates somewhat larger binding energies (per
unit reduced mass) at the ISCO than for the test particle case, for a
few moderate spin configurations. On the other hand, post-Newtonian
calculations (e.g. Blanchet 2002) suggest somewhat lower binding
energies per unit reduced mass at the ISCO (A. Buonanno; private
communication). Based on these results and on the limit $\mu
\rightarrow m$ for test particles (when $m \rightarrow 0$), we have
chosen to express inspiral losses in units of the reduced mass, $\mu$,
with magnitudes identical to those of the test particle cases,
irrespective of the BH mass combinations encountered in our models
(see Eq.~[2] and Table~\ref{tab:one}).

Losses to GWs during the plunge phase are much less well understood. An
exact result for the combined plunge + ringdown phase exists for the
test particle case, in the absence of any spin or orbital angular
momentum (Davis et al. 1971): it amounts to a loss of $\sim 0.01 M
q^2c^2$. We adopt this minimal loss for the plunge phase in our
slow-spin models. Orbital angular momentum should always be important
in astrophysical BH coalescences and it is likely that plunge losses
will then become substantially larger. For definiteness, we adopt ten
times larger losses during plunge for the fast-spin models (see, e.g.,
Nakamura, Oohara \& Kojima 1987). In the absence of analytical results
on the plunge phase for equal-mass binaries, we further assume that
the above test particle $q^2$ mass scaling is valid for any BH mass
combination (see Eq.~[3]).  Finally, we add a contribution to GW
losses from the ringdown phase. Our prescription is adapted from the
results of Khanna et al. (1999; extrapolated at large spin values),
with the same assumed $q^2$ mass scaling as before (see
Eq.~[4]).\footnote{A mass scaling with the ``reduced mass ratio,''
$\eta = \mu/(m+M)$, replacing $q$ in Eqs.~(3) and~(4) may be more
accurate in the limit $q \rightarrow 1$, according to post-Newtonian
calculations (S. Hughes, private communication). This would reduce the
importance of plunge and ringdown losses in our models, since $\eta
\rightarrow q$ when $q \rightarrow 0$ but $\eta \rightarrow 1/4$ when
$q \rightarrow 1$.}

\section{Results}

The evolution of the distributions of BH and galaxy masses in our
evolutionary models has been discussed extensively in paper I. We use
Monte-Carlo realizations of the merger tree of dark matter halos,
starting with $N \simeq 4.6 \times 10^4$ halos at $z=3$ with masses in
the range $10^{8.6}$--$10^{12.1}$~M$_\odot$.  Our merger tree database
effectively describes a fixed comoving volume $\sim 1.7 \times
10^4$~Mpc$^{3}$. It is then straightforward to calculate the comoving
mass density in BHs, $\rho_{\rm BH}$, and to monitor its evolution: we
simply follow the merger history of BHs and subtract, at each merger
event, the mass--energy lost to GWs. In models without any GW losses,
$\rho_{\rm BH} \simeq 1.4 \times 10^5$ M$_\odot$~Mpc$^{-3}$ and it
does not evolve with cosmic times. In models including GW losses,
however, a small fraction of $\rho_{\rm BH}$ is lost each time two BHs
coalesce, leading to a growing cumulative deficit.

\begin{figure}
\plotone{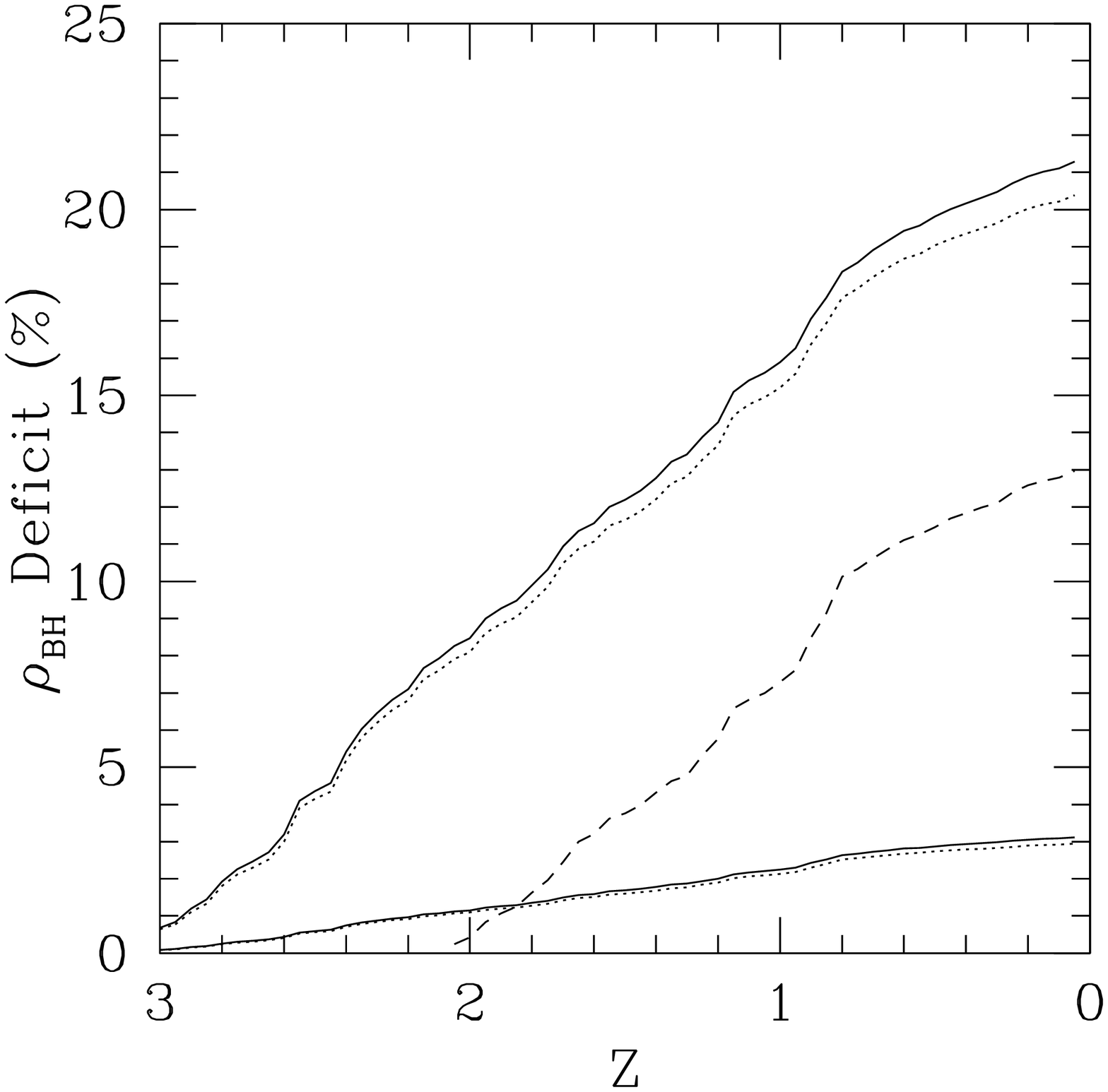}
\caption{Evolution with redshift, $z$, of the deficit in black hole
mass density, $\rho_{\rm BH}$, due to gravitational wave losses (as
compared to a model without any loss). The upper solid line
corresponds to the fast-spin model and the lower solid line to the
slow-spin model (see Table~\ref{tab:one}). Associated dotted lines
show results when the population of massive BHs is initially ten times
rarer and confined to the $10 \%$ most massive galaxies at $z =
3$. The dashed line corresponds to a fast-spin model starting at
$z=2$.
\label{fig:one}}
\end{figure}

Figure~\ref{fig:one} shows the evolution of the deficit in $\rho_{\rm
BH}$ from $z=3$ to $z=0$ in the slow- and fast-spin models of
Table~\ref{tab:one}. The cumulative deficit is relatively small in the
slow-spin model ($\sim 3 \%$ of the initial $\rho_{\rm BH}$ value) but
it reaches $\sim 20 \%$ in the fast spin-model. Models with a ten
times rarer population of massive BHs (initially confined to the $10
\%$ most massive galaxies) give essentially identical results (dotted
lines in Fig.~\ref{fig:one}). This is because most of the BH mass loss
occurs at the largest masses (see paper I).  The cumulative deficit
does depend on the value of the redshift at which cosmological
evolution is initiated, however, as shown by the dashed line in
Figure~\ref{fig:one} (a fast-spin model starting at $z=2$). This is a
consequence of the reduced total number of BH mergers.  We also note
that the cumulative $\rho_{\rm BH}$ deficits shown in
Figure~\ref{fig:one} are slightly smaller than the corresponding
values for the simpler slow- and fast-spin models discussed in paper
I. This results simply from the more realistic GW loss prescription
adopted here.

\begin{figure}
\plotone{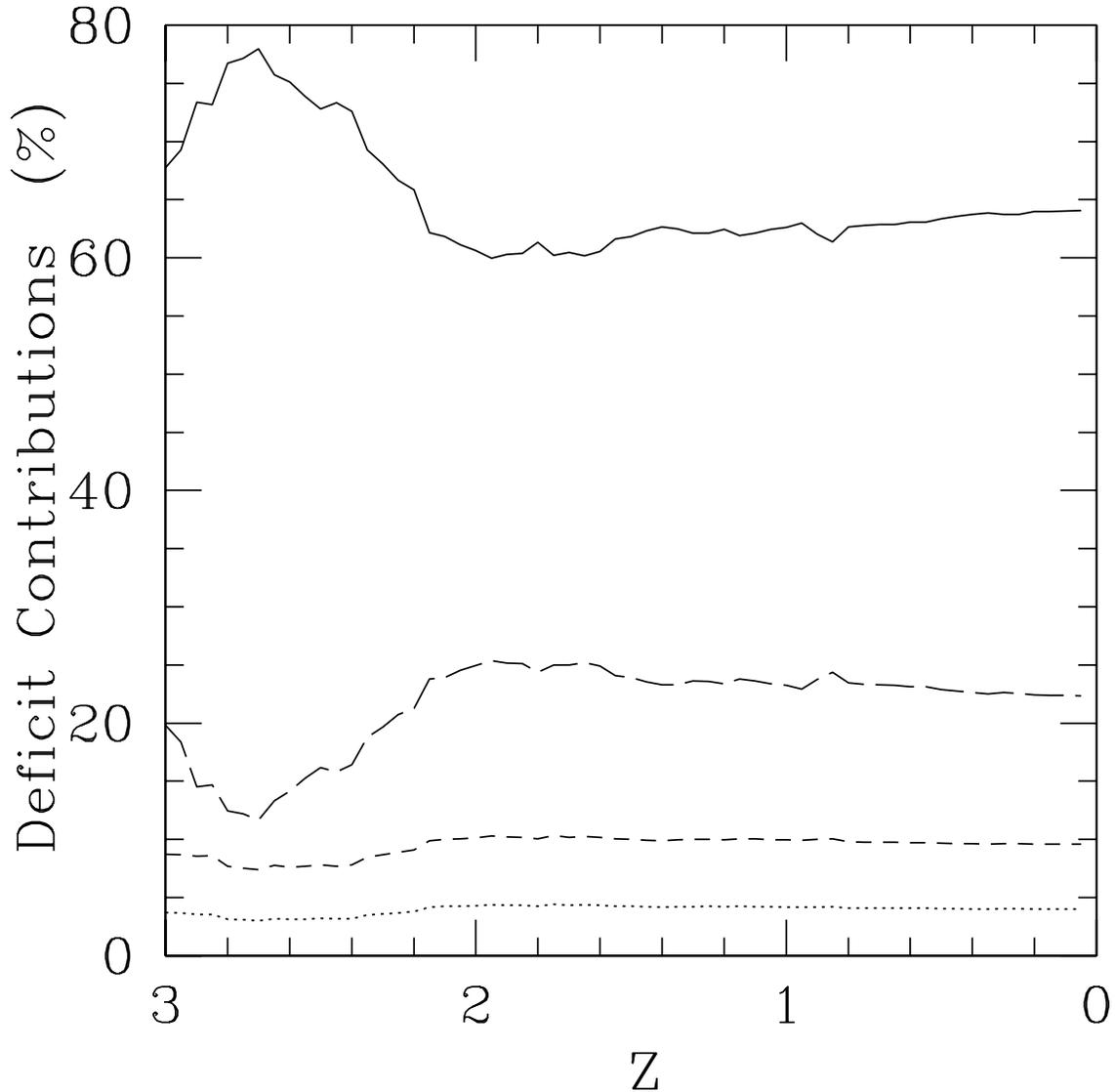}
\caption{Relative contributions to the deficit in black hole mass
density ($\rho_{\rm BH}$) from various sub-categories of gravitational
wave losses, as a function of redshift $z$, in the fast-spin
model. The solid line traces the dominant contribution from inspiral
losses of BH binaries with mass ratios $q < 0.5$, the long-dashed line
corresponds to inspiral losses from binaries with $q \geq 0.5$, the
short-dashed line to plunge losses from all binaries and the dotted
line to ringdown losses from all binaries. A qualitatively similar
loss hierarchy is obtained in the slow-spin model, except for a
negligibly small contribution from ringdown losses.
\label{fig:two}}
\end{figure}

Figure~\ref{fig:two} shows, for the fast-spin model, how various
sub-categories of GW losses contribute to the $\rho_{\rm BH}$
deficit. Little evolution with redshift is seen except early on, when
the initial BH masses are redistributed in galaxies. Inspiral losses
largely dominate the overall mass loss budget, while plunge and
ringdown losses contribute little. A combination of the adopted BH
masses and of the cosmological merger history experienced by BHs
results in most of the inspiral mass loss being due to BH binaries
with mass ratios $q < 0.5$ (compare solid and long-dashed lines in
Fig.~\ref{fig:two}; see also Fig.~5 in Menou 2003 for distributions of
BH mass ratios comparable to those found in our models). This is
important because inspiral losses, in the limit $q \ll 1$, are the
best known of all. Since the low $q$ limit is still relatively
accurate up to mass ratios $q \lsim 0.5$ according to post-Newtonian
calculations (see, e.g., discussion in Hughes \& Blandford 2003), this
indicates that our results may not be critically sensitive to various
uncertainties affecting our GW loss prescriptions for the other
regimes (see \S 2.2). Cumulative losses at $z=0$ correspond to
fractions $\sim \alpha_{\rm ins}/2$ in both the slow- and fast-spin
models (compare Fig.~\ref{fig:one} and Table~\ref{tab:one}). This
shows that a substantial fraction of the final mass density has been
assembled through mergers of BH binaries with $q < 0.5$. The exact
contribution to the mass assembly is difficult to estimate from losses
alone, however, because our prescription for inspiral losses (written
in units of reduced mass in Eq.~[2]) effectively reduces the losses
per unit ``real'' mass for large mass ratios ($\mu \rightarrow m/2$ in
the limit $q \rightarrow 1$).

\section{Discussion and Conclusion}

Cumulative mass loss to GWs during repeated cosmological BH
coalescences from $z \simeq 3$ to $z=0$ could reduce the amount of BH
mass locked into nearby dead quasars by up to $\sim$ 20 percent,
according to our fast-spin model (Fig.~\ref{fig:one}). This reduction
in the local BH mass density would effectively lead to a similar
fractional increase in the value of the radiative efficiency for
cosmic BH accretion, $\epsilon$, if a comparison between dead and
active quasars is attempted without accounting for GW losses.  Each
individual SMBH experiences numerous repeated mergers in its assembly
history (especially BHs in the most massive halos). However, our
detailed study of the merger history shows that the majority of these
mergers have small mass ratios, for which losses to GWs are equally
small (see Eqs.~[2-4]; note that the fraction of mergers with $q \sim
1$ can be significantly higher at $z\gsim 6$, where the effective
slope of the power spectrum at the mass--scale of collapsing objects
is shallower; Haiman 2004).

It is important to emphasize that our models are highly idealized and
that a number of effects ignored in our calculation are likely to
mitigate the already small magnitude of the $\rho_{\rm BH}$ deficit.
Except for the role of inefficient dynamical friction, we have assumed
maximally efficient BH coalescences and have thus maximized GW losses
in our models. The ``last parsec'' problem and gravitational radiation
recoil effects (Milosavljevic \& Merritt 2003; Favata, Hughes \& Holtz
2004), for example, will only make BH coalescences less frequent than
assumed here.\footnote{{Note that, by displacing or ejecting BHs
from galactic centers (e.g. Madau \& Quataert 2004), gravitational
radiation recoil could also lead to an underestimate of the mass
density in quasar remnants.}} We have also neglected the role of
orbital configurations in our loss prescriptions. For randomly
oriented BH encounters, some will be retrograde spin-orbit
configurations and lead to smaller inspiral losses than assumed in our
slow-spin model, even when BHs are spinning fast (e.g., $\alpha_{\rm
ins} \simeq 0.04$ for a maximally rotating retrograde configuration in
the test particle limit; see also Kojima \& Nakamura 1984). A more
accurate calculation should therefore account for the distribution of
orbital parameters of coalescing BHs and would probably find losses
intermediate between those predicted by our slow- and fast-spin
models. A proper calculation should also account for the growth of
$\rho_{\rm BH}$ with cosmic time due to accretion. We have effectively
maximized fractional losses by assuming that a given mass density is
in place at $z=3$ and that losses occur after this redshift without
any subsequent increase in $\rho_{\rm BH}$. Finally, typical BH spins
may be moderated by coalescences and accretion (Hughes \& Blandford
2003; Gammie et al. 2004), and this could easily bring losses closer
to the predictions of our slow-spin model.

Another model uncertainty is the limited range of masses described by
our merger tree. Losses in our models are dominated by the few most
massive BHs that happen to be present in our Monte-Carlo realizations
of the cosmological merger tree. These massive BHs still have lower
masses than the $> 10^8$~M$_\odot$ BHs of interest when comparing dead
and active quasars (see paper I and, e.g., Yu \& Tremaine 2002). We
have argued in paper I that including more massive BHs would increase
the value of the cumulative $\rho_{\rm BH}$ deficit, but this increase
is likely to remain modest. For example, a simple extrapolation with
BH mass of the $\rho_{\rm BH}$ deficit value predicted at $z=0$ shows
a small ($\ll \times 2$) increase of the fractional loss (shown in
Fig.~1) up to BH masses $\sim 10^9$~M$_\odot$.

Given the above arguments, the magnitude of $\rho_{\rm BH}$ deficits
shown in Figure~\ref{fig:one} cannot be taken at face value and it
appears likely that the losses amounting to $\sim 10$--$20 \%$ of
$\rho_{\rm BH}$ are only conservative upper limits to more realistic
values. The corresponding bias on the value of the radiative
efficiency of cosmic BH accretion, $\epsilon$, would also be $\lsim
10$--$20 \%$ and thus well within errorbars of current estimates
(e.g. Aller \& Richstone 2002; Elvis et al. 2002; Yu \& Tremaine
2002). Therefore, inferences that $\epsilon \gsim 0.1$ appear robust
and may indeed indicate radiatively--efficient accretion onto fast
spinning BHs. In the future, it it likely that the space
interferometer {\it LISA} will offer us some of the best empirical
constraints on the dark side of quasar evolution. Even though the
typical BH masses probed by {\it LISA} are smaller than those of
luminous quasars (e.g. Hughes 2002; Menou 2003), measurements of the
cosmological rate of massive BH coalescences and constraints on the
masses, and perhaps the spins, of these BHs will prove very useful to
clarify many of the uncertainties we have highlighted above.  A pulsar
timing array may also put interesting constraints on the magnitude of
the stochastic GW background generated by cosmological BH mergers
(e.g. Jaffe \& Backer 2003).

\section*{Acknowledgments}

K.M. thanks Alessandra Buonanno and Scott Hughes for helpful
discussions on general relativistic calculations, as well as the
Department of Astronomy at the University of Virginia for their
hospitality. Z.H. was supported in part by NSF through grants
AST-0307200 and AST-0307291.

\clearpage

\end{document}